\documentclass[reprint,
superscriptaddress,
nofootinbib,
amsmath,amssymb,
aps,
floatfix
]{revtex4-2}

\usepackage{graphicx}
\usepackage{dcolumn}
\usepackage{bm}
\usepackage{float}
\usepackage[switch*]{lineno}
\usepackage{xcolor}
\usepackage{adjustbox}
\usepackage{ctable}
\usepackage{multirow}
\usepackage[bottom]{footmisc}

\graphicspath{{Figures/}}

\begin{document}

\preprint{APS/123-QED}

\title{Variational inference of effective range parameters for \protect{${}^3$He-${}^4$He} scattering}

\author{Andrius Burnelis}
	\email{ab351021@ohio.edu}
    \affiliation{Department of Physics and Astronomy and Institute of Nuclear and Particle Physics, Ohio University, Athens, Ohio 45701, USA}

\author{Vojta Kejzlar}
    \email{vkejzlar@skidmore.edu}
    \affiliation{Department of Mathematics and Statistics, Skidmore College, Saratoga Springs, New York 12866, USA}

\author{Daniel Phillips}
    \email{phillid1@ohio.edu}
    \affiliation{Department of Physics and Astronomy and Institute of Nuclear and Particle Physics, Ohio University, Athens, Ohio 45701, USA}
    \affiliation{Department of Physics, Chalmers University of Technology, SE-41296 G\"oteborg, Sweden}

\begin{abstract}
    We use two different methods, Monte Carlo sampling and variational inference (VI), to perform a Bayesian calibration of the effective-range parameters in ${}^3$He-${}^4$He elastic scattering. The parameters are calibrated to data from a recent set of $^{3}$He-${}^4$He elastic scattering differential cross section measurements. Analysis of these data for $E_{\rm lab} \leq 4.3$ MeV yields a unimodal posterior for which both methods obtain the same structure. However, the effective-range expansion amplitude does not account for the $7/2^-$ state of ${}^7$Be so, even after calibration, the description of data at the upper end of this energy range is poor. The data up to $E_{\rm lab}=2.6$ MeV can be well described, but calibration to this lower-energy subset of the data yields a bimodal posterior. After adapting VI to treat such a multi-modal posterior we find good agreement between the VI results and those obtained with parallel-tempered Monte Carlo sampling.
\end{abstract}

\maketitle

\section{Introduction}

Two-body scattering calculations are ubiquitous in nuclear physics. Very often their output is compared with experimental data, and the parameters of the two-body interaction inferred, so the scattering can be extended to a different energy regime or embedded in another context: three-body scattering, fusion reactions, etc. 

$\chi^2$ minimization has been the workhorse for this inference for many years. However, a (naive) $\chi^2$ metric is only valid if the uncertainties on the data points are uncorrelated, and theoretical uncertainties do not contribute significantly to the problem's error budget. Strictly speaking, $\chi^2$ minimization also fails to provide information about parameter uncertainties, although the shape of the $\chi^2$ surface around the minimum can be examined to mitigate this issue. Bayesian methods have recently been used to address all these issues with scattering-parameter inference based on $\chi^2$-minimization~\cite{Pruitt:2024, Wesolowski:2018, Svensson:2021,  Odell:2022}. These methods have been used in different scattering formalisms: optical potentials, nucleon-nucleon scattering in Chiral Effective Field Theory, and R-matrix.

The multi-dimensional posterior probability density function (pdf) of the scattering parameters is typically examined through Markov chain Monte Carlo (MCMC) sampling. However, this can be computationally demanding. The computational requirements can be significantly reduced by the use of scattering-calculation emulators of either the black-box or intrusive variety~\cite{surmise2023,Melendez:2022kid,Drischler:2022ipa, Odell:2023cun}. In this paper we explore a complementary pathway, based on variational inference (VI)~\cite{Kucukelbir2017, Blei_2017, KejzlarVI23}.

In VI it is assumed that the posterior pdf takes a particular functional form, i.e., is part of a ``family'' of posteriors. The parameters of the pdf are computed by minimizing an objective function that is a surrogate for the distance of the variational form from the true posterior. VI is often conducted with a mean-field family, which assumes an uncorrelated product of distributions describes the model parameters' pdfs. A mean-field family of Gaussians was recently used to infer the pdfs of liquid-drop-model parameters from nuclear-mass data~\cite{Kejzlar:2020vla}. In this, or indeed any study that uses the mean-field Gaussian family, the parameters that specify the family are the means and variances of the uncorrelated Gaussians. But correlations between model parameters are also important, so here we extend the mean-field family to a family that parameterizes a general covariance matrix between the parameters of our scattering problem. The variational calculation now seeks to minimize the objective function in a $2 d$ + $d(d-1)/2$-dimensional space, but, as we shall see, variational inference is still markedly faster than MCMC sampling. 

As a test case for the efficacy of VI we selected a halo effective field theory (EFT) fit to $^3$He-$^4$He scattering data. The scattering amplitude in this case is straightforwardly related to the estimated parameters, since they are just the s- and p-wave effective-range-theory parameters. But it is critical that we can extract them reliably from data on elastic scattering, as they are key inputs to calculations of the capture reaction $^3$He($^4$He,$\gamma$)$^7$Be. 

This reaction produces $^{7}$Be in the Sun, after which it undergoes either a proton or electron capture reaction, both of which yield solar neutrinos at the upper end of the solar neutrino energy distribution~\cite{Adelberger2011}.Understanding the low energy elastic scattering of $^{3}$He and $^{4}$He thus has an important impact on the understanding of the solar pp-II and pp-III chains.

Solar fusion occurs in energy regimes where it is appropriate to use nuclei as the degrees of freedom. This motivates the use of a systematic EFT based on ${}^3$He and ${}^4$He degrees of freedom. In Ref.~\cite{Poudel_2022} halo EFT was used to extract low-energy ERPs from $^{3}$He-$^{4}$He scattering data~\cite{Poudel_2022}. These low energy parameters then have implications for solar fusion and neutrino production rates. Crucially, EFT comes with its own uncertainty quantification, since EFT calculations are carried out up to a finite order in the EFT expansion, and the error associated with omitted terms can be estimated. The EFT parameters may thus be extracted in a fit that includes the impact of model uncertainty on the inference.

In this paper we benchmark results using the same EFT, data, and MCMC methodology as employed in  Ref.~\cite{Poudel_2022} against VI results. Section~\ref{sec:VI} explains what variational inference is and how it works. Section~\ref{sec:theoryanddata} introduces the data of Ref.~\cite{Paneru} that we use for our fits, and also reviews the aspects of halo EFT that are needed to construct the likelihoods and priors we employ. Those likelihoods and priors, suitably combined, produce the Bayesian posterior of the EFT parameters, as we explain in Sec.~\ref{sec:Bayes}. In Sec.~\ref{sec:first_results} we perform MCMC sampling for an 
EFT-parameter posterior conditioned on the data of Ref.~\cite{Paneru} over a majority of the range of that experiment ($0.7 \leq E_{\rm lab} \leq 4.3$ MeV). However, the EFT, if not augmented by at least one resonance state, cannot describe the upper half of this energy range. We therefore restrict our fit to $E_{\rm lab} \leq 2.6$ MeV, and discover that the posterior is multi-modal. This necessitates the use of parallel tempering in our Markov Chain Monte Carlo. Our VI approach also requires an extension in order to produce multi-modal posteriors. We describe these extensions of our methods and display and compare the posteriors they produce in Sec.~\ref{sec:second_results}. We then provide a summary of our results in Sec.~\ref{sec:summary}. An appendix explains other, less effective, ways to deal with multi-modality in this posterior. 

\section{Variational Inference}\label{sec:VI}

Variational inference is an optimization-based approach to approximating posterior distributions. It is thus an alternative to the sampling-based MCMC approaches. The main idea is to approximate the target posterior distribution with a member of a simpler, more tractable  family of distributions. 

The first step is to posit a family of distributions $\mathcal{Q}$ indexed by variational parameters $\boldsymbol{\lambda}$. The family should be flexible enough to capture the shape (and potential correlations) of the target posterior distribution, but simple enough that the optimization process is efficient~\cite{Blei_2017}. 

The family member that best approximates the target distribution is then determined using an optimization process. The optimal member $q^{*}$ of $\mathcal{Q}$ is determined by minimizing the Kullback-Leibler (KL) divergence as a function of $\boldsymbol{\lambda}$. The KL divergence is defined as 
\begin{align}
    \label{eq:KL_div}
    \text{KL}(q(\boldsymbol{\theta}|\boldsymbol{\lambda})||p(\boldsymbol{\theta} | D)) = & \mathbb{E}_q \left[ \ln q(\boldsymbol{\theta}|\boldsymbol{\lambda})\right] \nonumber \\  & - \mathbb{E}_q \left[ \ln p(\boldsymbol{\theta},D) \right] + \ln p(D),
\end{align}
where $p(\boldsymbol{\theta}|D)$ is the target posterior distribution of unknown parameters $\boldsymbol{\theta}$ given the data $D$ and $\mathbb{E}_q$ denotes an expected value with respect to the density $q$.

In practice, the marginal data likelihood $p(D)$, the model evidence, is not a computationally tractable quantity. Instead, one maximizes an equivalent objective function called the evidence lower bound (ELBO):
\begin{equation}
    \label{eq:ELBO}
    \text{ELBO}(\boldsymbol{\lambda}) = \mathbb{E}_q \left[ \ln p(\boldsymbol{\theta}, D) \right] - \mathbb{E}_q \left[ \ln q(\boldsymbol{\theta}|\boldsymbol{\lambda}) \right].
\end{equation}
Since Eq.~\eqref{eq:ELBO} is equivalent to Eq.~\eqref{eq:KL_div} up to  an overall minus sign and a term that does not depend on $\boldsymbol{\lambda}$, the optimal distribution $q^{*}$ that maximizes the ELBO minimizes the KL divergence~\cite{Blei_2017}. In this way, we restate the posterior approximation problem as an optimization problem. In principle, ELBO can be minimized using any optimization approach, however, it is common practice to use stochastic gradient ascent (SGA) algorithm for its speed and scalability. SGA updates the variational parameter $\boldsymbol{\lambda}$ at the $t^{th}$ step according to
\begin{equation}\label{eqn:StochasticUpdate}
    \boldsymbol{\lambda}_{t+1} \leftarrow \boldsymbol{\lambda}_{t} + \rho_t \hat{\mathcal{L}}(\boldsymbol{\lambda}_t),
\end{equation}
where $\hat{\mathcal{L}}(\boldsymbol{\lambda})$ is an unbiased estimate of the ELBO gradient. Such an estimate can be readily obtained using  autodifferentiation as long as both the joint distribution $p(\boldsymbol{\theta}, D)$ and $q(\boldsymbol{\theta}|\boldsymbol{\lambda})$ are differentiable in $\boldsymbol{\theta}$. For $\hat{\mathcal{L}}(\boldsymbol{\lambda})$, a typical practice is to construct a simple Monte Carlo estimator using the samples from a variational distribution~\cite{Kucukelbir2017}.

\subsection{Choice of variational family}

The most popular variational family, the mean-field variational family, assumes that the unknown model parameters $\boldsymbol{\theta}= (\theta_{1},\dots,\theta_{m})$ are mutually independent. A general member of this family thus takes the form 
\begin{equation}
    \label{eq:mean_field}
    q(\mathbf{\theta}|\boldsymbol{\lambda}) = \prod_{j = 1}^{m}q_{j}(\theta_{j}|\boldsymbol{\lambda}_j).
\end{equation}
The functional form of each individual $q_{j}$ is up to the practitioner, however, the form of $q_{j}(\theta_{j})$ will affect the optimization efficiency and the resulting fidelity of the posterior approximation. Typical choices are the independent Gaussian variational family for real-valued parameters and the log-normal or Gamma variational family for non-negative parameters. With the right choice of particular $q_{j}$'s, the posterior's means should agree well with the true posterior. However, the major flaw of the mean-field family is that it assumes a decoupled covariance structure, i.e., zero correlation between the parameters being inferred. The resulting variational approximation underestimates the uncertainties in the case of correlated parameters~\cite{Blei_2017}. 

A simple remedy to this flaw of a mean-field family is to posit a full rank Gaussian variational family:
\begin{equation}
    \label{eq:full_rank}
    q(\boldsymbol{\theta}|\boldsymbol{\lambda}) = \mathcal{N}(\boldsymbol{\theta}|\boldsymbol{\mu}, \boldsymbol{\Sigma}),
\end{equation}
where $\boldsymbol{\lambda}$ is now$(\boldsymbol{\mu}, \boldsymbol{\Sigma})$, i.e., it includes the mean vector and a positive definite covariance matrix.

Variational families are an active area of research and many other flexible variational families exist. We refer the reader to the work of~\cite{pmlr-v130-ambrogioni21a} for a detailed discussion on variational families and their implementation.

\subsection{Implementation details}

To guarantee the positive-definiteness of the covariance matrix $\boldsymbol{\Sigma}$ for the Gaussian variational family~\eqref{eq:full_rank}, we parameterize the covariance matrix in terms of its Cholesky decomposition as a product of a lower triangular matrix with positive diagonal entries $\boldsymbol{L}$ and its transpose. Additionally, all the strictly positive variational parameters $\lambda$ were transformed as
\begin{equation}\label{eqn:reparametrization}
    \Tilde{\lambda} = \log (e^{\lambda} - 1)
\end{equation}
to avoid constrained optimization.

When it comes to the practicalities of ELBO minimization via SGA~\eqref{eqn:StochasticUpdate}, choosing an optimal learning rate $\rho_t$ can be challenging. Ideally, the rate should be low when the Monte Carlo estimates of the ELBO gradient are unstable (high variance) and high when the estimates are stable (low variance). The elements of the variational parameter $\lambda$ can also vary significantly in scale, requiring the learning rate to accommodate these varying, often small, scales. The rise of stochastic optimization in machine learning has spurred the creation of numerous algorithms that provide element-wise adaptive learning rates. We use the Adaptive Moment Estimation (Adam) algorithm~\cite{Adam}, which is known for its popularity and ease of implementation.

\section{Data and Scattering Model}\label{sec:theoryanddata}

\subsection{Experimental data}

\begin{figure*}
    \includegraphics[scale = 0.5]{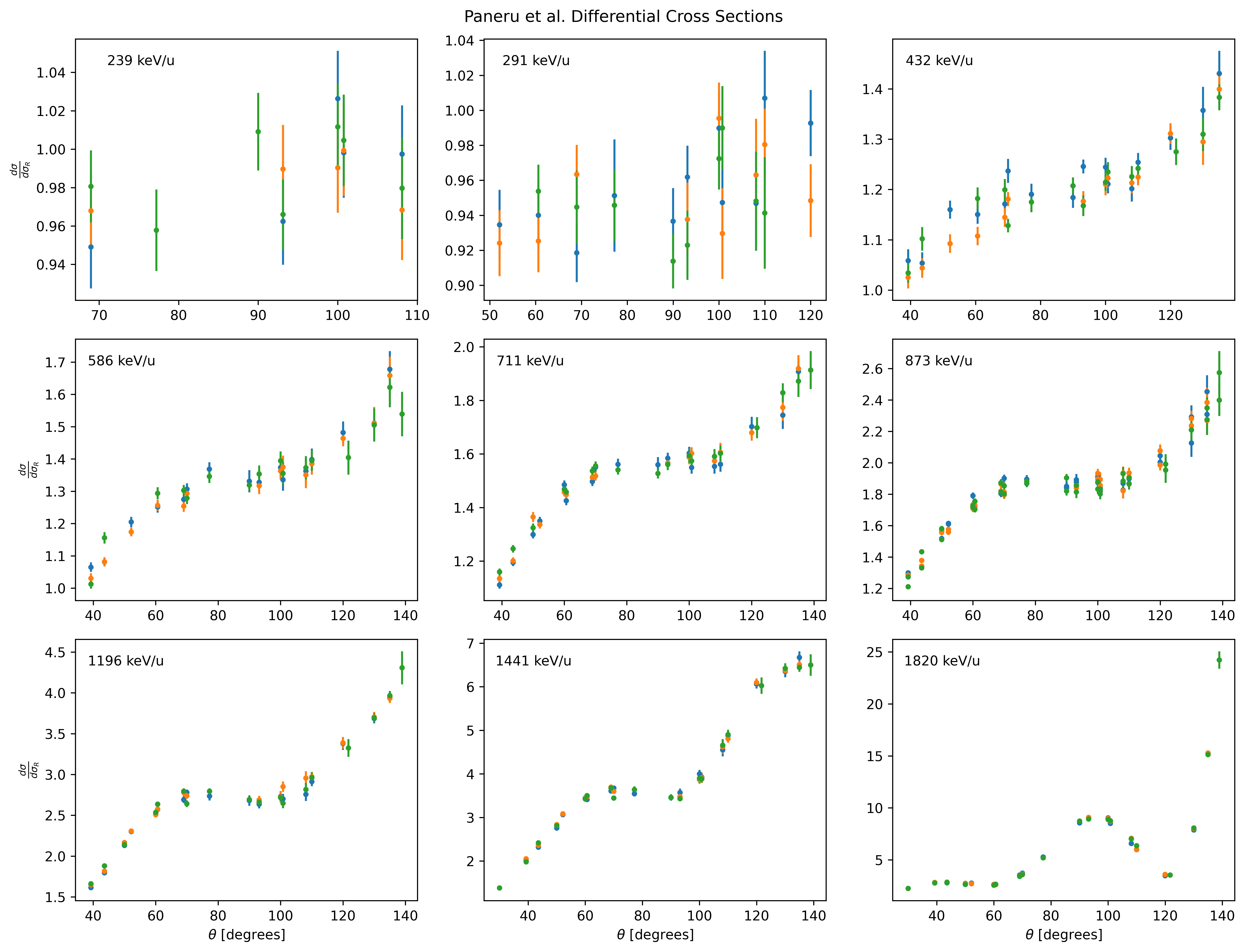}
    \caption{\label{fig:dcs_data}Differential cross section data of Paneru et al. relative to the Rutherford cross section versus detector angle. Each panel corresponds to the beam energy indicated in it~\cite{Paneru}.}
\end{figure*}

The data used in this analysis comes from the elastic $^3$He-$^4$He scattering experiment performed at TRIUMF by Paneru {\it et al.}~\cite{Paneru}. This experiment impinged a $^{3}$He beam, with an energy ranging from 0.7 to 5.5 MeV, on a ${}^4$He gas target contained in the SONIK target cell. The detector array covered an angular range of $30^{\circ} < \theta_{CM} < 139^{\circ}$. Table~\ref{tab:norm_priors} lists the number of different angles at which data was obtained for each of the ten beam energies chosen (two runs used $E_{\rm lab}=2.624$ MeV). It also specifies the (fractional) common-mode error (i.e., the normalization uncertainty) in the cross section for each energy. 
The data values and their point-to-point uncertainties are represented in Fig.~\ref{fig:dcs_data}. Full details can be found in Ref.~\cite{Paneru}.

The quantity $N_{data}$ represents the total number of data included in the analysis. Table~\ref{tab:norm_priors} contains the number of measurements at each energy. We perform analysis on two different subsets of the data; 0.706-2.624 MeV and 0.706-4.342 MeV. The values of $N_{data}$ are 293 and 398 respectively. $N_{E}$ represents the number of energy bins (panels shown in Fig.~\ref{fig:dcs_data}) we are including in the analysis. $N_{\theta}^{E}$ represents the number of angles within a particular energy bin.

\subsection{Parameterization of scattering amplitude}

Halo EFT for this problem reproduces the modified effective-range expansion~\cite{Bethe_1949,Hammer:2017,Poudel_2022}. The parameters we seek to estimate are thus the effective-range parameters (ERPs) for low-energy $^{3}$He-$^{4}$He elastic scattering. 

The differential cross section is given by 
\begin{equation}
    \frac{d \sigma}{d \Omega} = |f_{c}|^{2} + |f_{i}|^{2}.
\end{equation}
The non-spin-flip amplitude $f_c$ includes a ``Rutherford amplitude'' that represents pure-Coulomb scattering of the two nuclei. The 
rest of $f_c$, as well as the spin-flip amplitude $f_i$, can be expanded as a sum of partial waves \cite{Spiger_1967}. We thus have:
\begin{align}
    f_{c} = - \frac{\eta}{2 k} \csc^{2}(\theta / 2) & \exp\left( i \eta \log(\csc^{2}(\theta / 2)) \right) \nonumber \\ + \frac{1}{k} \sum_{\ell = 0}^{\infty} & \exp(2 i \alpha_{\ell})  P_{\ell}(\cos \theta) \nonumber \\ & \times \left[ \frac{\ell + 1}{\cot \delta_{\ell}^{+} - i} + \frac{\ell}{\cot \delta_{\ell}^{-} - i} \right] 
    \label{eq:fc}
\end{align}
\begin{align}
    f_{i} = \frac{1}{k} \sum_{\ell = 0}^{\infty} & \exp(2 i \alpha_{\ell}) \sin \theta \frac{d P_{\ell}(\cos \theta)}{d \cos \theta} \nonumber \\
    \times & \left[ \frac{1}{\cot \delta_{\ell}^{-} - i} - \frac{1}{\cot \delta_{\ell}^{+} - i}\right].
    \label{eq:fi}
\end{align}
Here $\alpha_{\ell}$ is the difference of Coulomb phase shifts, $\alpha_\ell \equiv \sigma_l - \sigma_0$, $\theta$ is the scattering angle in the center of mass frame, $\delta_{\ell}^{\pm}$  the $\ell$th phase shift of the $\pm$ scattering channel, and $P_{\ell}$ is the $\ell$th Legendre polynomial. In this analysis, we only consider s- and p-wave amplitudes and so truncate the sums in Eqs.~(\ref{eq:fc}) and (\ref{eq:fi}) at $\ell=1$. 

Hamilton, Overb\"o and Tromberg~\cite{Hamilton:1973} showed that the effective range function is real, analytic, and holomorphic in the physical energy sheet, cut along the negative real axis. It is also regular at the origin ($k = 0$) so there is a well defined Taylor expansion at the corresponding energy~\cite{Sparenberg:2009}. They also showed how to modify this expansion to account for long ranged interactions such as the Coulomb interaction. We make use of the modified effective range expansion (ERE) which lets us calculate the phase shifts via:
\begin{align}
\label{eq:invamp}
    k^{2 \ell + 1} ( \cot \delta_{\ell}^{\pm} - i) = \frac{2 k_{c}}{e^{-\pi \eta}} \Bigg[ & \frac{\Gamma{(\ell + 1)}^{2} k_{c}^{2 \ell} K_{\ell}^{\pm}(k)}{(\ell^{2} + \eta^{2}) |\Gamma(\ell + i \eta)|^{2}} \nonumber \\ 
    & - \frac{k^{2 \ell} H(\eta)}{|\Gamma(1 + i\eta)|^{2}} \Bigg]
\end{align}
with 
\begin{eqnarray}
	H(\eta) = \Psi(i \eta) + \frac{1}{2 i \eta} - \ln(i \eta).
\end{eqnarray}
and $\eta = \frac{z_{1} z_{2} \alpha \mu}{\hbar k}$ the Sommerfeld parameter.

Our model parameters come from the effective range function $K_{\ell}^{\pm}(k)$ given as
\begin{equation}
    K_{\ell}^{\pm} = \frac{1}{2 k_{c}^{2 \ell + 1}} \left[-\frac{1}{a_{\ell}^{\pm}} + \frac{1}{2} r_{\ell}^{\pm} k^{2} + \frac{1}{4} P_{\ell}^{\pm} k^{4} + \mathcal{O}(k^{6}) \right].
\end{equation}
The coefficients of the powers of $k^{2}$ are the effective range parameters (ERPs). The ERPs of interest for this analysis are $\frac{1}{a_{0}}$, $r_{0}$, $\frac{1}{a_{1}^{\pm}}$, $r_{1}^{\pm}$, and $P_{1}^{\pm}$. The ERE amplitude must have a pole at the momenta corresponding to the bound state energies $B = k_{bs}^{2} / 2\mu$, and so the right-hand side of Eq.~(\ref{eq:invamp}) is zero when analytically continued to $k= i k_{bs}$. There are two shallow p-wave bound states in the $\frac{3}{2}^{-}$ and the $\frac{1}{2}^{-}$ channels (1.5866 and 0.43 MeV respectively)\cite{Tilley2002}. This constraint reduces the number of sampling parameters, since it relates $\frac{1}{a_{1}^{\pm}}$ to the bound state momenta $k_{bs}^{\pm}$, and the other ERPs of that channel~\cite{Poudel_2022}. Similarly, we can relate $r_{1}^{\pm}$ and $P_{1}^{\pm}$ to the asymptotic normalization coefficients $(C_{1}^{\pm})^{2}$ through the derivative evaluated at the bound state energy~\cite{Poudel_2022}.

Halo EFT is a systematic expansion of the scattering amplitude in powers of $Q \equiv \max\{p, q\} / \Lambda_{B}$. 
In Halo EFT the different contributions of the effective-range theory amplitude are organized in a hierarchy of importance. 
The convergence pattern of the EFT is consistent with the ordering presented in~\cite{Poudel_2022}. In what follows we work with the EFT amplitude at next-to-next-to-leading order, NNLO, $O(Q^2)$. Theory uncertainties are thus of a relative size $\sim Q^3$. Their impact on the posterior pdf can be included through a theory covariance matrix, as will be described further in Sec.~\ref{sec:Bayes}.

One effect that causes the breakdown of the EFT is the existence of an f-wave resonance at 5.22 MeV in the lab frame~\cite{Tilley2002}. We have not included the physics of this resonance in our analysis. For this reason, we do not expect the EFT to make accurate predictions for energies approaching 5.22 MeV. 

We also model the portion of the systematic uncertainty of the experimental data that is associated with uncertainties in the beam current and target density. These uncertainties are completely correlated across all the data taken at a particular energy; they are a ``common-mode error''. We account for them by introducing additional parameters $f_i$ that multiply the theoretical prediction at energy $E_i$~\cite{DAgostini:1993, Odell:2022}. The $f_i$'s are assigned (Gaussian) priors based on the expected size of this common-mode error, as reported in Ref.~\cite{Paneru}, and repeated in the third row of Table~\ref{tab:norm_priors}.

\section{Bayesian PDFs}\label{sec:Bayes}

We adopt a shorthand notation to represent the sampled parameters. The vector $\boldsymbol{\theta} = (\boldsymbol{a}, \boldsymbol{f})$ where $\boldsymbol{a}$ is a $1 \times 6$ dimensional vector of ERPs and $\boldsymbol{f}$ is a $1 \times N_{E}$ dimensional vector of normalization coefficients. $N_{E}$ is the number of energy bins considered in the analysis and $N_{\text{data}}^{E}$ is the number of data points in the energy bin $E$. The joint posterior distribution we wish to obtain is
\begin{equation}
    p(\boldsymbol{\theta} | D, I) = \frac{p(D | \boldsymbol{\theta}, I) p(\boldsymbol{\theta} | I)}{p(D | I)} \propto p(D | \boldsymbol{\theta}, I) p(\boldsymbol{\theta} | I).
\end{equation}
Here $p(D | \boldsymbol{\theta}, I)$ is the likelihood function and $p(\boldsymbol{\theta} | I)$ is the prior on the parameters $\boldsymbol{\theta}$. $p(D | I)$ is the model evidence, also sometimes called the ``marginal likelihood''. In practice, this piece of the right-hand side of Bayes' theorem functions as a normalization, and does not affect the shape of the pdf for the parameters $\boldsymbol{\theta}$. 

We define the likelihood function  via
\begin{equation}
    p(D | \boldsymbol{\theta}, I) = \frac{1}{\sqrt{{(2 \pi)}^{N_{\rm data}} \det{\Sigma}}} \exp \left( - \frac{\chi^{2}}{2} \right)
\end{equation}
where
\begin{eqnarray}
    \chi^{2} = \sum_{j, k = 1}^{N_{E}} \sum_{l, m = 1}^{N_{\text{data}}^{E}} \left( y_{j, l} - f_{j} y(E_{j}, \theta_{l}, \boldsymbol{a}) \right)  [\Sigma^{-1}]_{jl, km} \nonumber \\ \times \left( y_{k, m} - f_{k} y(E_{k}, \theta_{m}, \boldsymbol{a}) \right).
    \label{eq:chisq}
\end{eqnarray}

We consider two different choices for the covariance matrix $\Sigma$.
\begin{itemize}
\item A version $\Sigma \equiv \Sigma^{\text {exp}}$ that accounts only for the point-to-point experimental uncertainties:
\begin{equation}
    \label{eq:exp_unc}
    \Sigma_{jl, km}^{\text{exp}} = \sigma_{jl}^{2} \delta_{jl, km}.
\end{equation}

\item A covariance matrix which includes an additional term, that accounts for theory uncertainties. Reference~\cite{Wesolowski:2018} showed that the next-order (in this case $O(Q^3)$) pieces of the EFT amplitude can be marginalized over. In the simplest case this yields a theory covariance matrix of the form:
\begin{equation}
    \label{eq:th_unc}
    \Sigma^{\text{th}}_{jl, km} = {(y_{\text{ref}})}_{jl} {(y_{\text{ref}})}_{km} \bar{c}^{2} Q_{jl}^{3} Q_{km}^{3},
\end{equation}
with $Q$ the EFT expansion parameter and $\bar{c}$ the rms value of the EFT expansion coefficients. In what follows we take the hyperparameters $Q$ and $\bar{c}^{2}$ that appear in $\Sigma^{\text th}$ as fixed, basing the adopted values, $\Lambda_{B} = 1.014 fm^{-1}$ and $\bar{c} = 0.7$, on the analysis of the convergence pattern of the EFT carried out in Ref.~\cite{Poudel_2022}. Once theory uncertainties are included  the full covariance matrix is  $\Sigma \equiv \Sigma^{\text{th}} + \Sigma^{\text {exp}}$.
\end{itemize}

Turning our attention now to the prior $p(\boldsymbol{\theta} | I)$,
this pdf has two distinct pieces. The first pertains to the effective-range theory parameters, and the second is related to the normalization parameters. We take these to be independent prior pdfs:
\begin{equation}
    \label{eq:prior_decomp}
    p(\boldsymbol{\theta} | I) = p(\boldsymbol{a} | I) p(\boldsymbol{f} | I) .
\end{equation}

The priors for the effective range parameters are taken to be truncated normal distributions
\begin{equation}
    \label{eq:a_prior}
p(\boldsymbol{a} | I) = \prod_{i}^{d} \mathcal{N}(\mu_{i}, \sigma_{i}^{2}) T(a_{i}, b_{i}),
\end{equation}
with
\begin{equation}
    T(a, b) = \begin{cases}
        1 & [a, b] \\
        0 & \text{otherwise}.
    \end{cases}
\end{equation}
These truncated normal distributions encode the notion that the ERPs $r_0$ and $P_1^{\pm}$ are natural when measured in units of the breakdown scale $\Lambda_{B}$. The ERT parameters corresponding to terms of positive mass dimension in the amplitude are not natural, and have enhancements or suppressions, as detailed in Ref.~\cite{Poudel_2022}. This too is captured by the priors. The hyperparameters of the prior pdf for the ERT parameters are given in Tab.~\ref{tab:param_priors}. 

Meanwhile, the prior for the normalization parameters $f_{i}$ are taken to be
\begin{equation}
    \label{eq:f_prior}
 p(\boldsymbol{f} | I) = \prod_{i}^{g} \mathcal{N}(1, \sigma_{f_{i}}^{2}) T(0, 2).
\end{equation}
Here the $\sigma_{f_i}^{2}$s are the variances of associated with the normalization uncertainty at each energy that was reported in Ref.~\cite{Paneru}. They are listed in Tab.~\ref{tab:norm_priors}. 

\begin{table*}[ht]
    \caption{\label{tab:norm_priors}Number of data points in each energy bin and the variances for each of the Gaussian priors of the normalization coefficients. See Eq.~\ref{eq:f_prior} for the full prior.}
    \centering
    \begin{tabular}{p{0.095\linewidth}p{0.08\linewidth}p{0.08\linewidth}p{0.08\linewidth}p{0.08\linewidth}p{0.08\linewidth}p{0.08\linewidth}p{0.08\linewidth}p{0.08\linewidth}p{0.08\linewidth}p{0.08\linewidth}p{0.08\linewidth}}
        \toprule
        $E_{lab}$ (MeV) & 0.706 & 0.868 & 1.292 & 1.759 & 2.137 & 2.624 & 2.624 & 3.598 & 4.342 & 5.484 \\
        \toprule
        $N_{data}^{E}$ & 17 & 29 & 45 & 46 & 52 & 52 & 52 & 52 & 53 & 53 \\
        $\sigma_{f_{i}}$ & 0.064 & 0.076 & 0.098 & 0.057 & 0.045 & 0.062 & 0.041 & 0.077 & 0.063 & 0.089 \\
        \toprule
    \end{tabular}
\end{table*}

\begin{table*}[ht]
    \caption{\label{tab:param_priors}Truncation bounds, means, and variances of the Gaussian priors for each of the effective range parameters. Equation~(\ref{eq:a_prior}) shows the full prior on the effective range parameters.}
    \centering
    \begin{tabular}{p{0.143\linewidth}p{0.12\linewidth}p{0.12\linewidth}p{0.12\linewidth}p{0.12\linewidth}p{0.12\linewidth}p{0.12\linewidth}}
        \toprule
        Parameter & $\frac{1}{a_{0}}$ & $r_{0}$ & ${(C_{1}^{+})}^{2}$ & $P_{1}^{+}$ & ${(C_{1}^{-})}^{2}$ & $P_{1}^{-}$ \\
        \toprule
        $a$ & -0.02 & -3.0 & 5.0 & -6.0 & 5.0 & -6.0 \\
        $b$ & 0.06 & 3.0 & 25.0 & 6.0 & 25.0 & 6.0 \\
        $\mu$ & 0.025 & 0.8 & 13.84 & 0.0 & 12.59 & 0.0 \\
        $\sigma^{2}$ & 0.015 & 0.4 & 1.63 & 1.6 & 1.85 & 1.6 \\
        \toprule
    \end{tabular}
\end{table*}

\section{Results}\label{sec:results}

In this section we compute four different posterior pdfs for the ERPs and the normalization parameters, using both VI and various MCMC techniques. Once analysis is completed, a comparison of posterior distributions will indicate the level of agreement between the two methods.

The choice of MCMC methods has two motivations: the high dimensionality of the problem, and the desire to obtain a full pdf for the model parameters. Once the log-posterior function (or log-prior and log-likelihood functions separately in the case of \texttt{ptemcee}) is coded, MCMC is easily implemented using the prebuilt python packages \texttt{emcee}\cite{Foreman_Mackey_2013} and \texttt{ptemcee}\cite{Vousden_2015}.

We performed four versions of the analysis. The first uses only the experimental covariance matrix and analyzes the Paneru {\it et al.}~\cite{Paneru} data up to $E_{\rm lab}=4.342$ MeV. It is described in Subsec.~\ref{sec:first_results}. We also present a version of this analysis where we evaluated the $\chi^2$ (\ref{eq:chisq}), and hence the likelihood, using the full covariance matrix $\Sigma^{\text{exp}} + \Sigma^{\text{th}}$.
In Subsec.~\ref{sec:second_results} we analyze
data over a smaller energy range, where the EFT we used to obtain the scattering amplitude is reliable. This yields a bimodal posterior pdf for the ERPs and so necessitates some modification of our approach to variational inference of the posterior. In parallel to the results of Subsec.~\ref{sec:first_results}, Subsec.~\ref{sec:second_results} reports both an analysis of the low-energy portion of the data of Ref.~\cite{Paneru} that uses just the experimental covariance matrix as well as one where we combine the experimental uncertainties with the theoretical ones by taking $\Sigma \equiv \Sigma^{\text{exp}} + \Sigma^{\text{th}}$.

\subsection{Unimodal results using higher-energy data}\label{sec:first_results}

In this subsection we present results from analysis of the SONIK dataset with energies ranging from $0.7$ to $4.3$ MeV. At first we do not consider the theory uncertainties. 

Table~\ref{tab:chi2dof} shows the $\chi^{2}/{\rm dof}$ of the model at the maximum a posteriori (MAP) value solution for our parameters from~\cite{Poudel_2022}, compared to the scattering data. For this we have used $\chi^2$ with both $\Sigma \equiv \Sigma^{\text{exp}}$ and $\Sigma \equiv \Sigma^{\text{exp}} + \Sigma^{\text{th}}$. As expected, given that the EFT does not include the resonance at $E_{\rm lab}=5.22$ MeV, the $\chi^2$ becomes large as $E_{\rm lab}$ approaches that energy. 

\begin{table}[h]
    \centering
    \caption{\label{tab:chi2dof}The $\chi^{2}$ per degree of freedom of the model at the MAP value solution for our parameters from~\cite{Poudel_2022}. The reported values were computed using Eq.~\ref{eq:chisq} with the appropriate covariance matrix.}
    \begin{tabular}{p{0.25\linewidth}p{0.3\linewidth}p{0.3\linewidth}}
        \toprule
        $E_{\text{max}}$ (MeV) & $\chi^{2}/\text{dof}$ & $\chi^{2}/\text{dof}$ \\ 
         & ($\Sigma \equiv \Sigma^{\text{exp}}$) & ($\Sigma \equiv \Sigma^{\text{exp}} + \Sigma^{\text{th}}$) \\
        \toprule
        0.706 & 1.2320 & 1.1925 \\
        0.868 & 1.2306 & 1.1863 \\
        1.292 & 1.7751 & 1.6773 \\
        1.759 & 1.9463 & 1.8748 \\
        2.137 & 2.1392 & 2.1234 \\
        2.624 & 2.3837 & 2.3410 \\
        3.598 & 4.8966 & 3.4682 \\
        4.342 & 19.9067 & 10.4666 \\
        \bottomrule
    \end{tabular}
\end{table}
        
The physics model is therefore very unlikely to be correct throughout the entire energy regime. However, the problem of posterior computation still has a unique answer. We can use that answer to assess the usefulness of VI for posterior computation in a situation where we have a sizable amount of data. We emphasize, though, that the posteriors of the estimated parameters discussed in this subsection do not reflect a realistic result for the physics model under consideration. They are presented as the outcome of a technical exercise in comparing MCMC parameter estimation to VI parameter estimation. 

We first use the Python package \texttt{emcee} to generate a Markov chain of samples that represents the posterior distribution of the ERPs and the normalization parameters. Our production run had $50,000$ burn-in steps, and ran for a total of $300,000$ sample steps, with $30$ walkers. To start the sampling, we draw random starting positions from the priors for each walker. The sampler evolves through the burn-in steps to reach equilibrium. Upon completing burn-in, the sampler then begins the sample acquisition process by storing chains of samples for each walker. The sampling method was \texttt{emcee}'s default stretch move and the autocorrelation time was 287. The sampling was done with a standard computer and took approximately 4 hours to complete the burn-in and sampling.

To generate the VI results we utilized the Adam optimizer from the \texttt{pytorch} library~\cite{Ansel:2024}. The learning rate was set to $5 \times 10^{-3}$, and the number of optimization steps was $30,000$. This method took less than 10 minutes to complete. Once we obtain the optimized variational parameters, we then draw samples from the variational family.

\begin{figure*}[t]
    \includegraphics[width = 0.75\textwidth]{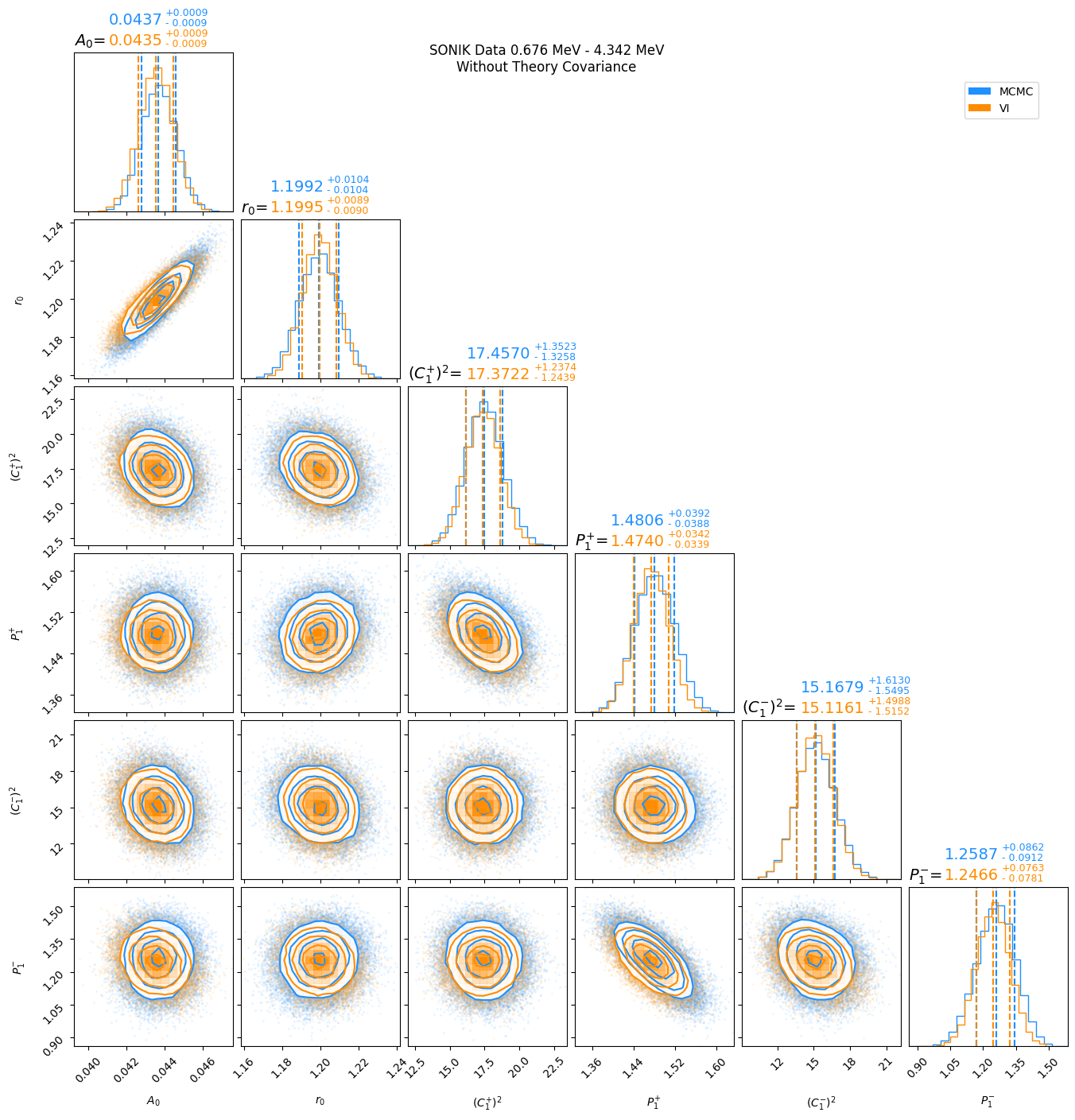}
    \caption{\label{fig:4.3_no_cov}Joint posterior density of ERPs from analyzing $E_{\rm lab}=0.676$ MeV to 4.342 MeV data without including the theory covariance. The blue distribution was generated by sampling with \texttt{emcee}, and the orange is the distribution from VI.}
\end{figure*}

The joint posterior density obtained using these two methods is shown in Fig.~\ref{fig:4.3_no_cov}. It is unimodal. The MCMC result is shown in blue and the VI result in orange. The two methods produce remarkably consistent results: Tab.~\ref{tab:percent_diff_43_n} compares the percent differences of the medians and standard deviations between the VI and the MCMC approach. The two-dimensional distributions, as quantified by the correlation coefficients, also agree very well. Perhaps this is not surprising, given that MCMC reveals this posterior to be well approximated by a multi-variate Gaussian, and this is the family we have used for the VI.\@ We remark in passing that the agreement is, predictably, not nearly as good if the ``mean-field'' (i.e.\ uncorrelated) VI family is employed here. 

\begin{table}[h]
    \centering
    \caption{\label{tab:percent_diff_43_n}The percent differences ($ \equiv |x_{1} - x_{2}| / \bar{x} * 100\%$) of the $68\%$ equal tail credible intervals obtained for the ERPs using VI and {\tt ptemcee} without including the theory covariance in the likelihood.}
    \begin{tabular}{p{0.18\linewidth}p{0.25\linewidth}p{0.25\linewidth}p{0.25\linewidth}}
        \toprule
        Parameter & 16\% Quantile & 50\% Quantile & 84\% Quantile \\
        \toprule
        $A_{0}$ & 0.35\% & 0.34\% & 0.36\% \\
        $r_{0}$ & 0.14\% & 0.02\% & 0.10\% \\
        $(C_{1}^{+})^{2}$ & 0.02\% & 0.49\% & 1.07\% \\
        $P_{1}^{+}$ & 0.12\% & 0.45\% & 0.76\% \\
        $(C_{1}^{-})^{2}$ & 0.13\% & 0.34\% & 0.99\% \\
        $P_{1}^{-}$ & 0.08\% & 0.97\% & 1.65\% \\
        \toprule
    \end{tabular}
\end{table}

The second calculation using the data set up to $E_{\rm lab}=4.3$ MeV included the theory covariance. It had the same number of burn-in, sample steps, and walkers. To include the theory covariance we use $\Sigma \equiv \Sigma^{\text{th}} + \Sigma^{\text {exp}}$ in Eq.~(\ref{eq:chisq}). The run with the theory covariance had an autocorrelation time of 383. This run also completed in approximately 4 hours.

\begin{figure*}[t]
    \includegraphics[width = 0.75\textwidth]{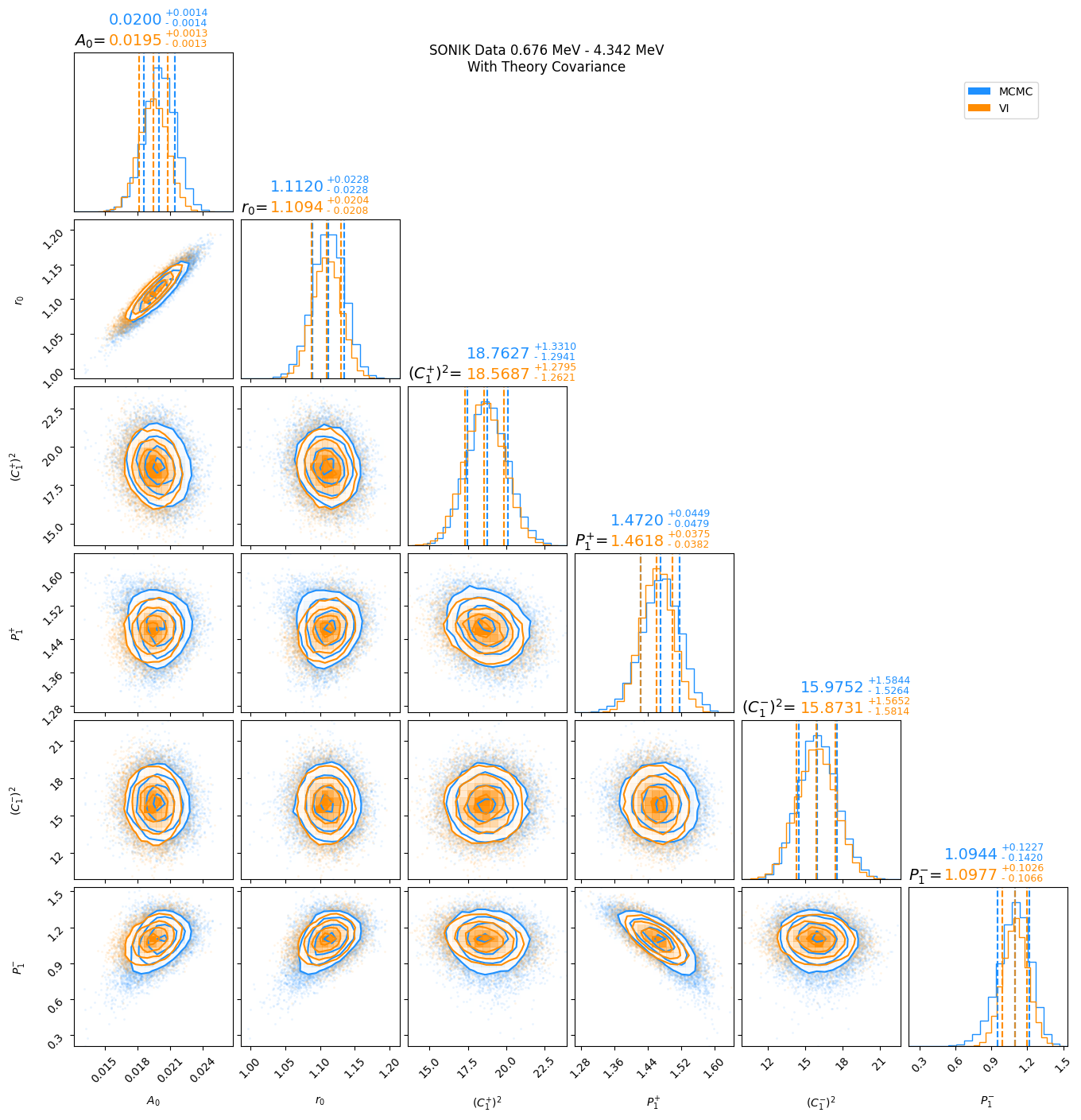}
    \caption{\label{fig:4.3_yes_cov}Joint posterior density of ERPs from analyzing $E_{\rm lab}=0.676$ MeV to 4.342 MeV data with the theory covariance. The blue distribution was generated by sampling with \texttt{emcee}, and the orange is the distribution from VI.}
\end{figure*}

This joint posterior density is shown in Fig.~\ref{fig:4.3_yes_cov}, with the same color scheme as in Fig.~\ref{fig:4.3_no_cov}. Once again VI and MCMC agree very well. This time the posteriors for the ERPs are broader, and the correlations are somewhat different to that found in the previous analysis. But the conclusion is that VI has no problem with the more complex correlation structure of the data that results when theory uncertainties are considered. 

\begin{table}[h]
    \centering
    \caption{\label{tab:percent_diff_43_y}Percent differences of the $68\%$ equal tail credible intervals obtained for the ERPs  using {\tt ptemcee} and VI when the theory covariance is included in the likelihood.}
    \begin{tabular}{p{0.18\linewidth}p{0.25\linewidth}p{0.25\linewidth}p{0.25\linewidth}}
        \toprule
        Parameter & 16\% Quantile & 50\% Quantile & 84\% Quantile \\
        \toprule
        $A_{0}$ & 2.42\% & 2.77\% & 3.09\% \\
        $r_{0}$ & 0.06\% & 0.23\% & 0.43\% \\
        $(C_{1}^{+})^{2}$ & 0.93\% & 1.04\% & 1.23\% \\
        $P_{1}^{+}$ & 0.03\% & 0.69\% & 1.17\% \\
        $(C_{1}^{-})^{2}$ & 1.09\% & 0.64\% & 0.69\% \\
        $P_{1}^{-}$ & 3.98\% & 0.31\% & 1.38\% \\
        \toprule
    \end{tabular}
\end{table}

\subsection{Bimodal results using only lower-energy data}\label{sec:second_results}

Preliminary analysis of cross section data from $E_{lab} = 0.7 \text{ to } 2.6$ MeV using \texttt{emcee} produces bimodal distributions, with the bimodality being driven by two possible solutions for the pair of shape parameters $\{P_{1}^{+},P_{1}^{-}\}$. This bimodality was also observed in the analysis of Poudel and Phillips~\cite{Poudel_2022}. It appears to result from the inability of lower-energy cross section data to distinguish the roles of the two p-wave channels that have different total angular momentum. This bimodality at best produces a very long autocorrelation time in {\tt emcee}, and at worst results in the MCMC sampling being poorly converged and unreliable.

In order to handle the bimodal distribution we modified our sampling approach and used the \texttt{ptemcee} sampler instead~\cite{Vousden_2015}.\ \texttt{ptemcee} uses the parallel tempering Monte Carlo technique which is better equipped to handle multimodal distributions~\cite{Gregory_2005}. Sample chains generated with \texttt{ptemcee} produced much lower autocorrelation times than \texttt{emcee} in this application. 

We mirror the analyses done in Sec.~\ref{sec:first_results} and do two runs, one with and one without the theory covariance. In the case where the likelihood did not include the theory covariance matrix, the autocorrelation time decreased from $287$ to $2$ while when we use the likelihood including theory covariance it decreased from $383$ to $2$. Because \texttt{ptemcee} produces chains with much lower autocorrelation times, we do not need as many steps as we employed with \texttt{emcee}.

\begin{figure*}
    \includegraphics[width = 0.75\textwidth]{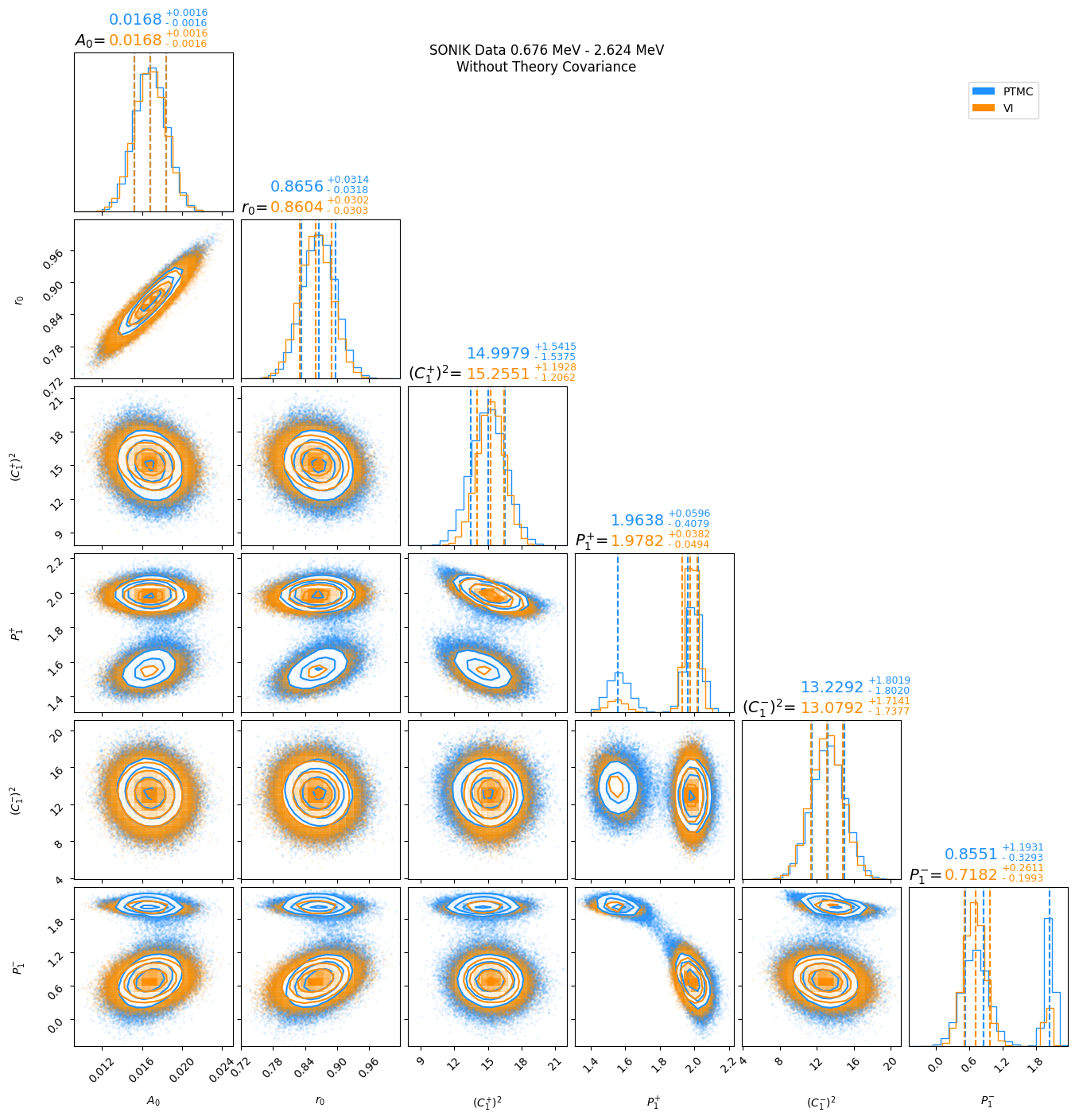}
    \caption{\label{fig:2.6_no_cov}Joint posterior density of ERPs from analyzing the data of Ref.~\cite{Paneru} from $E_{\rm lab}=0.676$ MeV to 2.624 MeV without the theory covariance using both \texttt{ptemcee} and VI. The blue distribution was generated by sampling with \texttt{ptemcee}, and the orange is the distribution from VI.}
\end{figure*}

\begin{figure*}
    \includegraphics[width = 0.75\textwidth]{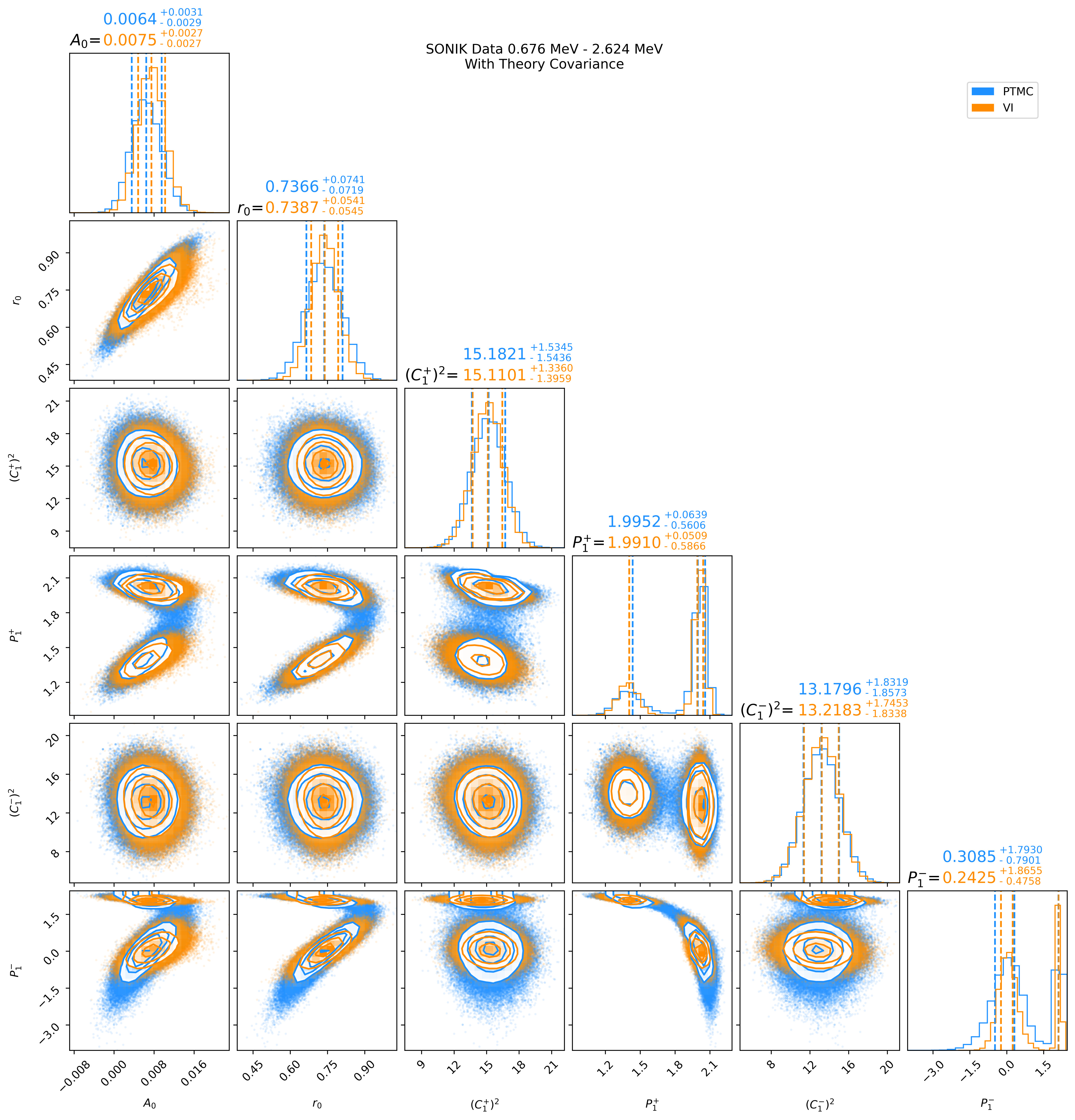}
    \caption{\label{fig:2.6_yes_cov}Joint posterior density of ERPs from analyzing the data of Ref.~\cite{Paneru} from $E_{\rm lab}=0.676$ MeV to 2.624 MeV with the theory covariance included. The blue distribution was generated by sampling with \texttt{ptemcee}, and the orange distribution from VI.}
\end{figure*}

For both \texttt{ptemcee} analyses we used $10,000$ burn-in steps, and $50,000$ sample steps and sampled across 8 different inverse temperatures with values of $\{1, 0.917, 0.841, 0.771, 0.707, 0.5, 0.353, 0.25 \}$. Figures~\ref{fig:2.6_no_cov} and~\ref{fig:2.6_yes_cov} show the joint posterior densities without and with the theory covariance respectively. 

As alluded to earlier in this section, the VI approach requires a modification to allow for multimodal posteriors as the full rank Gaussian variational family~\eqref{eq:full_rank} is not flexible enough to describe multimodality directly. The VI-based multimodal posterior distributions in Figs.~\ref{fig:2.6_no_cov} and~\ref{fig:2.6_yes_cov}  were obtained using the Black Box Variational Bayesian Model Averaging (BBVBMA) algorithm~\cite{Kejzlar:2023}. The BBVBMA posterior variational distribution is a mixture distribution, where each mixture component is a full rank Gaussian produced by the standard VI approach (as described in Section~\ref{sec:VI}) with random initialization. Each mixture weight is consequently proportional to the resulting ELBO value. The resulting multimodal approximations are  based on a mixture of 1000 VI approximations computed independently and in parallel on a computer cluster. We refer the reader to Ref.~\cite{Kejzlar:2023} for full details of BBVBMA.

The posteriors presented here are bimodal due to the two possible solutions for the pair of shape parameters $\{P_{1}^{+}, P_{1}^{-}\}$. These two parameters are anticorrelated as shown in the bottom row, fourth column of both Figs.~\ref{fig:2.6_no_cov} and~\ref{fig:2.6_yes_cov}. The results generated by \texttt{ptemcee} tend to be broader than those from VI. The distribution obtained through \texttt{ptemcee} provides samples bridging between the two modes, while VI does not. 

Tables~\ref{tab:percent_diff_26_n} and~\ref{tab:percent_diff_26_y} show the percent differences between the ERP quantiles obtained with \texttt{ptemcee} and VI without and with the theory covariance respectively. There are a few large discrepancies between the quantiles: the differences in inference in the two approaches are most noticeable for the parameters in which bimodality appears. In the analysis without the theory covariance, we see a percent difference of $21\%$ in the $16\%$ quantile of $P_{1}^{+}$, while in  the $84\%$ quantile for $P_{1}^{-}$ we have a $70\%$ difference. Meanwhile, the analysis with the theory covariance has percent differences of $-69\%$ and $23\%$ in the $16\%$ and $50\%$ quantiles of $P_{1}^{-}$, although the quantiles and univariate distribution for $P_1^+$ are in remarkable agreement across the methods in this case. 

In general the univariate distributions agree well between VI and \texttt{ptemcee} for parameters that do not exhibit bimodality. There are moderate differences in the $16\%$, $50\%$, and $84\%$ quantiles of $A_{0}$ but $A_0$'s median is a small number. We conclude that VI performs well for the unimodal parameter distributions even though the overall distribution is bimodal.

\begin{table}[h]
    \centering
    \caption{\label{tab:percent_diff_26_n}  Percent differences of the $68\%$ equal tail credible intervals for the ERPs without including the theory covariance for the $E_{\rm lab}=0.676$-- 2.624 MeV analysis.}
    \begin{tabular}{p{0.18\linewidth}p{0.25\linewidth}p{0.25\linewidth}p{0.25\linewidth}}
        \toprule
        Parameter & 16\% Quantile & 50\% Quantile & 84\% Quantile \\
        \toprule
        $A_{0}$ & 0.01\% & 0.10\% & 0.10\% \\
        $r_{0}$ & 0.46\% & 0.61\% & 0.73\% \\
        $(C_{1}^{+})^{2}$ & 4.26\% & 1.72\% & 0.53\% \\
        $P_{1}^{+}$ & 21.41\% & 0.73\% & 0.35\% \\
        $(C_{1}^{-})^{2}$ & 0.84\% & 1.17\% & 1.58\% \\
        $P_{1}^{-}$ & 1.18\% & 17.41\% & 70.63\% \\
        \toprule
    \end{tabular}
\end{table}

\begin{table}[h]
    \centering
    \caption{\label{tab:percent_diff_26_y} Percent differences of the $68\%$ equal tail credible intervals for the ERPs with the theory covariance included for the analysis up to $E_{\rm lab}=2.624$ MeV.}
    \begin{tabular}{p{0.18\linewidth}p{0.25\linewidth}p{0.25\linewidth}p{0.25\linewidth}}
        \toprule
        Parameter & 16\% Quantile & 50\% Quantile & 84\% Quantile \\
        \toprule
        $A_{0}$ & 30.75\% & 15.06\% & 7.04\% \\
        $r_{0}$ & 2.90\% & 0.28\% & 2.24\% \\
        $(C_{1}^{+})^{2}$ & 0.53\% & 0.46\% & 1.61\% \\
        $P_{1}^{+}$ & 2.17\% & 0.23\% & 0.84\% \\
        $(C_{1}^{-})^{2}$ & 0.60\% & 0.32\% & 0.28\% \\
        $P_{1}^{-}$ & -69.37\% & 23.75\% & 0.32\% \\
        \toprule
    \end{tabular}
\end{table}

\section{Summary}\label{sec:summary}
Inferring parameters of scattering models from large data sets is a common nuclear-physics problem. Here we have shown that Variational Inference (VI) can expedite this parameter estimation. The two sets of scattering data we examined in Sec.~\ref{sec:results} had 398 and 293 data points respectively, and the VI posterior was obtained in a fraction of the time needed for the sampling. We found essentially perfect agreement between the VI result for the posterior of the ${}^3$He-$\alpha$ effective-range expansion parameters and the posterior obtained using Monte Carlo sampling when the parameter posterior pdf was unimodal.

A critical ingredient of the success of the variational inference was that the VI family, i.e., the parameterization of the posterior, was flexible enough to allow for correlations. An accurate description of correlations between parameters is critical for reliable Uncertainty Quantification. 

VI requires that we assume a parameterization of the posterior. Here we used a multivariate Gaussian. Other families, e.g., a multi-variate $t$-distribution, can be implemented. 

Analyzing the subset of data from Ref.~\cite{Paneru} for $E_{\rm lab} \leq 2.6$ MeV produced a bimodal posterior. We used parallel tempered Monte Carlo Markov Chains in order to sample this posterior thoroughly; the autocorrelation time  for standard MCMC sampling was almost prohibitively long. Long autocorrelation times are often a signal of multi-modality. 

In the case of VI the multi-modality mainfested as a sensitivity of the solution to the Evidence Lower Bound Optimization (ELBO) to its starting position. We leveraged this feature of VI to obtain different optima for the ELBO, and then combined the inferred distributions using the ELBO estimate of the Bayesian evidence. This modified VI approach produced posteriors that were in good agreement with those found using parallel tempered MCMC. 

Source code for generating the \texttt{emcee} and \texttt{ptemcee} results is available at \url{https://github.com/AndriusBurn/3he_alpha_mc_sampling}. The source code used for generating the VI results is available at \url{https://github.com/kejzlarv/VBI_Pytorch_3HEscattering}.

\color{black}

\appendix
\section{Other ways to combine single-peak results}

While performing the analysis discussed in Subsec.~\ref{sec:second_results}, we investigated different methods where we could reliably combine two single-peak results in a way that preserved the relative peak heights and shapes of the two modes. In this appendix we discuss two alternative methods we considered as well as the caveats that go along with them. Both methods discussed here follow a model comparison strategy; we treat samples obtained from one peak ($\boldsymbol{\theta}_{A}$) model $A$, and the samples obtained from the other peak ($\boldsymbol{\theta}_{B}$) model $B$. The relative peak heights are thus determined by the ratio of the marginal likelihoods.

\subsection{Harmonic Mean Estimator}

The harmonic mean estimator was introduced in Ref.~\cite{Newton:1994} which showed how using samples from the posterior one could approximate the marginal likelihood. We can start by examining the quantity
\begin{equation}
    \label{eq:reciprocal_likelihood}
    \rho = \mathbb{E}_{P(\boldsymbol{\theta} | D, I)} \left[ \frac{1}{\mathcal{L}(\boldsymbol{\theta})} \right] \equiv \int \frac{P(\boldsymbol{\theta} | D)}{\mathcal{L}(\boldsymbol{\theta})} d \boldsymbol{\theta}
\end{equation}
where $\mathcal{L}$ is the likelihood function, and $\boldsymbol{\theta}$ is a sample drawn from the posterior. This is the expectation of the reciprocal likelihood with respect to the posterior. We can relate this to the marginal likelihood. Through applying Bayes' theorem to the posterior in Eq.~(\ref{eq:reciprocal_likelihood}), we have
\begin{equation} \label{eq:marginal_likelihood_relation}
    \rho = \int \frac{1}{\mathcal{L}(\boldsymbol{\theta})} \frac{\mathcal{L}(\boldsymbol{\theta}) P(\boldsymbol{\theta} | I)}{P(D | I)} d \boldsymbol{\theta} = \frac{1}{P(D | I)}.
\end{equation}
The harmonic mean estimator is given by
\begin{equation}
    \label{eq:hme}
    \hat{\rho} = \frac{1}{N}\sum_{i = 1}^{N} \frac{1}{\mathcal{L}(\theta_{i})}
\end{equation}
where $\mathcal{L}$ is the likelihood function, and $\theta_{i}$ is a sample drawn from the posterior. We can take our samples from each of the peaks and compute corresponding $\hat{\rho}_{A}$ and $\hat{\rho}_{B}$ and then take the ratio to obtain the relative height of the peaks. 

This method however yields inconsistent and often times incorrect results. We can recast understand its failure from the perspective of importance sampling. For importance sampling to be effective, we would need the sampling distribution to be broader than the target distribution. In this case, we are drawing from the posterior distribution which is narrow when compared to the prior. For this reason, we are not obtaining effective samples and this method fails.

\subsection{Overlap Normalization}

Suppose we have two peaks in a posterior distribution that have a slight overlap between them. We can then use a uniform prior to perform sampling for each peak separately. If we then construct the priors so there is a small overlap between them---one that coincides with the overlap between the peaks we can cross-normalize the two sets of samples. 

\begin{figure}[H]
    \centering
    \includegraphics[scale = 0.45]{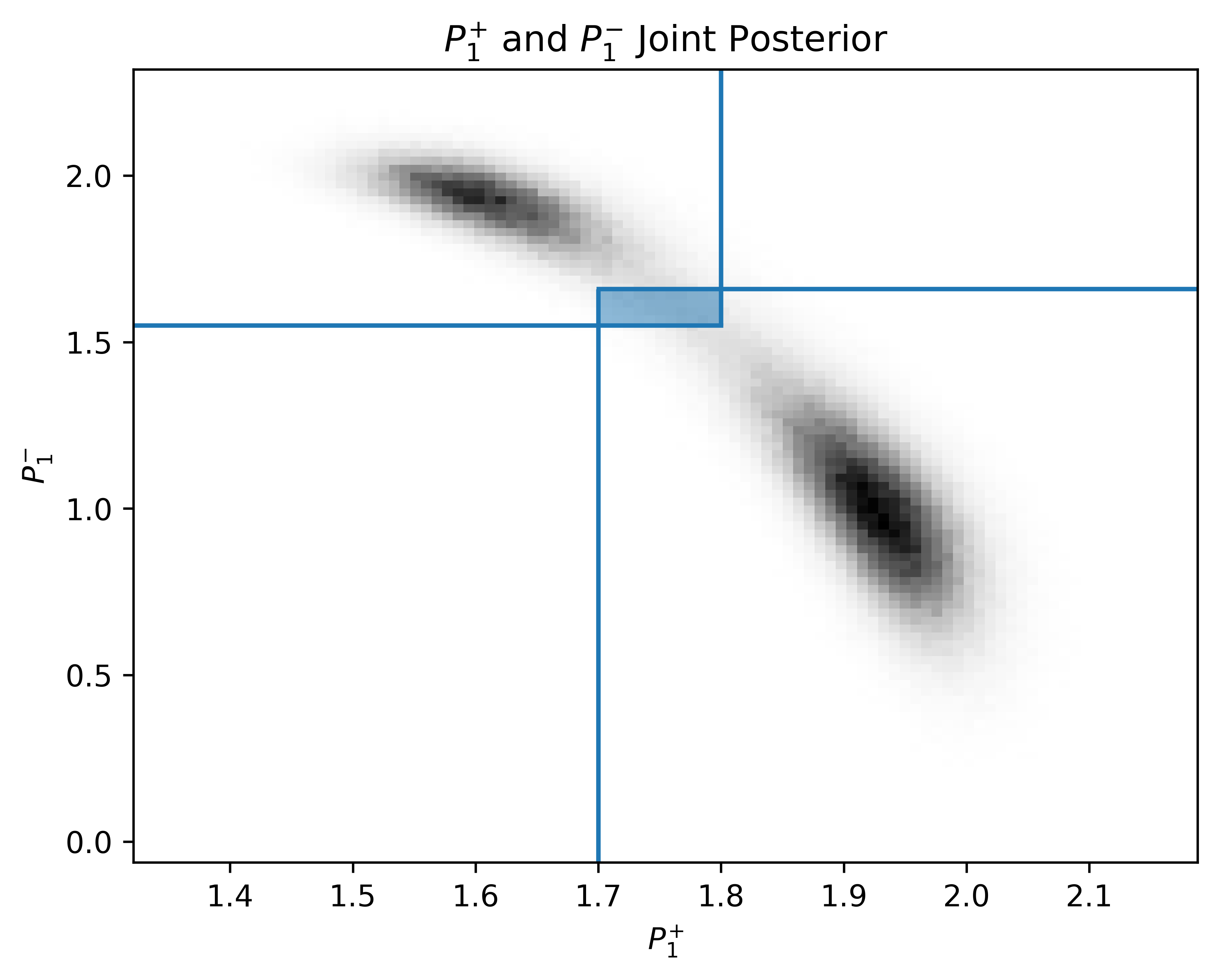}
    \caption{\label{fig:overlap_method}An illustration showing the overlapping region set by the two priors. The solid blue lines are the truncation cuts of the two priors, and the shaded region is the overlap region.}
\end{figure}

An illustration of this is shown in Fig.~\ref{fig:overlap_method}.
In the overlap region, the samples from the two runs are drawn from the same posterior distribution. We can exploit the uniqueness of the underlying distribution within this region to obtain the relative peak heights. 

Suppose there are $N_{A}$ total samples in set $\boldsymbol{\theta}_{A}$, and $N_{B}$ total samples in set $\boldsymbol{\theta}_{B}$, with $N_{A, ol}$ and $N_{B, ol}$ the number of samples appearing in the overlap region. Without loss of generality, we may assume $N_{A, ol} > N_{B, ol}$. We can determine a relative peak height ratio of
\begin{equation}
    \label{eq:overlap_peak_height}
    r = \frac{N_{B, ol}}{N_{A, ol}}.
\end{equation}
With this target ratio, we can then thin the appropriate set of samples such that after thinning we have
\begin{equation}
    \label{eq:overlap_thinning}
    \frac{N_{B,\text{thinned}}}{N_{A, \text{thinned}}} = r.
\end{equation}
Although both numbers of samples have been given the ``thinned'' subscript, it is sufficient to only thin one of the two sets of samples. Once the thinning is done and the ratio of the total thinned samples is the target ratio $r$, we can then combine the two sets of samples to obtain the full joint posterior distribution with the appropriate relative peak heights.

While this method can be useful, it is not always possible to have a clear overlap region. The overlap region can often be very small, and the number of samples in this region is often very low. This can lead to a very noisy estimate of the relative peak heights. Normalizing the peaks using this method yields posteriors that closely resemble the ones obtained from \texttt{ptemcee}. But, ultimately choosing {\tt ptemcee} proved more effective in our case.

\bibliographystyle{unsrt}
\bibliography{citations}

\end{document}